\documentclass[12pt]{article}
\usepackage{epsfig}

\def\beq{\begin{equation}}
\def\eeq#1{\label{#1}\end{equation}}
\def\eeqn{\end{equation}}

\def\beqa{\begin{eqnarray}}
\def\eeqa#1{\label{#1}\end{eqnarray}}
\def\eeqan{\end{eqnarray}}

\let\bar=\overbar

\def\etal{{\it et al.}}
\def\ie{{\it i.e.}}
\def\eg{{\it e.g.}}

\def\Dslash{\not{\hbox{\kern-4pt $D$}}}
\def\dslash{\not{\hbox{\kern-2pt $\del$}}}

\def\msb{{\bar{\ssstyle M \kern -1pt S}}}

\providecommand{\etal}{et~al.}              
\providecommand{\etc}{etc.}                 
\providecommand{\sun}{\odot}
\def\Title#1{\begin{center} {\Large {\bf #1} } \end{center}}

\begin{document}

\Title{The Optical Afterglows of Gamma-Ray Bursts}

\bigskip\bigskip


\begin{raggedright}  

{\it Stephen T. Holland\index{Holland, S. T.}\\
Department of Physics\\
University of Notre Dame\\
Notre Dame, IN 46556--5670, U.S.A.}
\bigskip\bigskip
\end{raggedright}


\section{Introduction}

     The first gamma-ray burst (GRB) for which an optical afterglow
was observed was GRB~970228~\cite{GGV1997,VGG1997}, and the first
redshift for a GRB came from the optical afterglow of
GRB~970508~\cite{MDS1997,MDK1997}.  The inferred redshift for
GRB~970508 was $z = 0.835$, which demonstrated that GRBs have
cosmological origins and required that the bursts have isotropic
equivalent energies of $10^{52}$--$10^{53}$ erg, which was difficult
to explain.  Over the next five years intensive efforts were made to
observe the optical afterglows of GRBs.  The ROTSE telescope saw
prompt optical emission from GRB~990123 just 22 seconds after the
burst~\cite{AM1999,ABB1999}, and~\cite{HH1999} identified a possible
star-forming region at the location of this burst.  The purpose of
this article is to provide a brief overview of optical observations of
GRBs and what they can tell us about the physics, geometry, and
environments of the bursts.


\section{The Relativistic Fireball Model}

     Over the past few years a consensus has developed that the
optical afterglow of a GRB is caused by an expanding relativistic
fireball that is independent of the details of the central engine.  In
this picture the optical emission is produced when the expanding
fireball collides with material surrounding the progenitor to produce
an external shock.  Electrons in the shocked material are accelerated
and acquire a power law distribution of energies, $N(\gamma_e) \propto
\gamma_e^{-p}$, for electrons with $\gamma_e$ greater than some
minimum value.  The electrons then emit synchrotron radiation which we
observe as the afterglow.  A detailed review of the expanding fireball
model and the production of optical afterglows is given by
\cite{M2002}.

     The optical flux from the afterglow, $f_\nu$, is related to the
frequency, $\nu$ and time since the burst, $t$, by $f_\nu(\nu,t)
\propto \nu^\beta t^\alpha$.  The indices $\alpha$ and $\beta$ are in
turn related to the electron index, $p$, in ways that depend on the
values of the cooling, synchrotron, and self-absorption frequencies
($\nu_c$, $\nu_m$, and $\nu_a$).  If both $\alpha$ and $\beta$ can be
determined observationally then this information can be used to
determine the relative ordering of these three characteristic
frequencies.  Once this is known the electron index and other physical
parameters of the fireball can be determined.


\section{The Spectral Energy Distribution}

\begin{figure}[htb]
   \begin{center}
      \epsfig{file=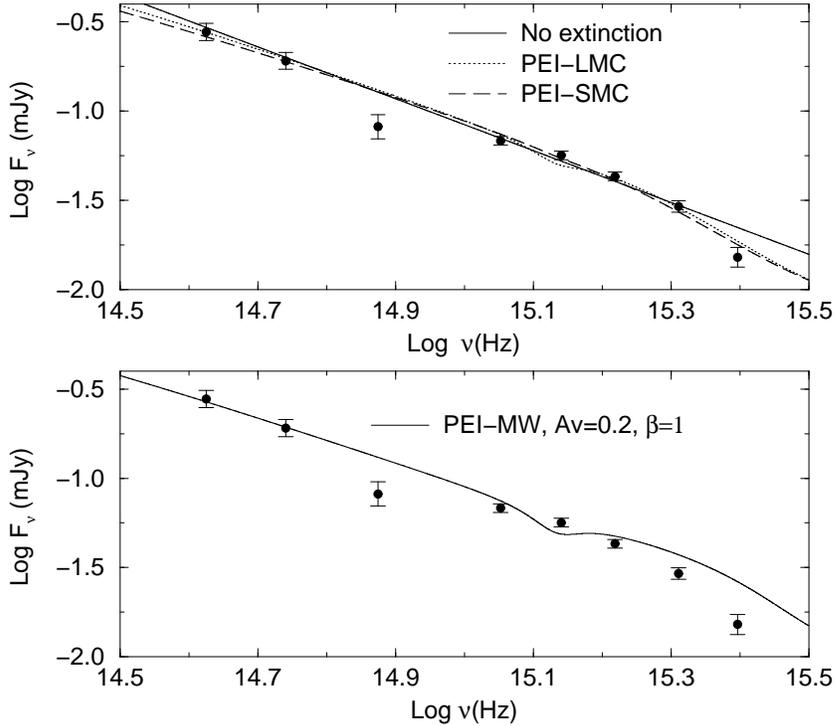,height=4.0in}
      \caption{The observed specific flux $F_{\nu}$ vs.\ the
               rest-frame frequency $\nu$ normalized to 2000
               Sept.~27.9~UT for GRB~000926 from the $K$ to $U$ bands.
               The upper panel shows that the observed curvature can
               be well fitted with $\beta=-1$ and a SMC or LMC
               extinction law with $A_V = 0.2$.  The lower panel shows
               that the MW extinction law is inconsistent with the
               data~\cite{FGD2001}.\label{FIGURE:grb000926}}
   \end{center}
\end{figure}

     The key to interpreting the spectrum is the fact that a
synchrotron spectrum is made up of power laws, except near the
spectral breaks.  Observed spectra however, often show curvature due
to extinction along the line of sight to the burst.  This curvature
can be used to estimate the amount of extinction in the host galaxy by
assuming that the observed spectrum is the result of an intrinsic
power law that has been reddened by some extinction law.  The
intrinsic spectral slope of an afterglow can be determined by
reddening an assumed intrinsic power law spectrum until it matches the
observed spectrum.  This allows both the intrinsic $\beta$ and the
amount of extinction in the host galaxy along the line of sight to the
burst to be determined (see Fig.~\ref{FIGURE:grb000926}).

     The relationship between the spectral slope and the electron
index depends on the relative ordering of the cooling and synchrotron
frequencies with respect to the optical, and there are cases where
$\beta$ is independent of $p$.  Some case can usually be ruled out by
the intrinsic value of the spectral slope.  Also, if spectra from
other frequency bands are available, such as $X$-ray data, then they
can also be used to estimate the value of the electron index.  The
requirement that the electron index must be the same at all
frequencies can help rule out some orderings of $\nu_c$ and $\nu_m$.

     Models for ten GRB afterglows are presented by~\cite{PK2002}.
The mean electron index for these bursts is $\overline{p} = 1.9$, but
five bursts have $p < 2$.  In the standard relativistic fireball model
$p < 2$ represents infinite energy in the electrons thus is
unphysical.  This problem can be avoided by introducing an upper limit
for the electron energy distribution, However, detailed modelling of
the acceleration of particles in highly relativistic shocks predict
that the electron index should be $\approx 2.3$~\cite{AGK2001}, which
is inconsistent with what is seen in many bursts.  The fact that many
GRBs appear to have electron indices of less than two may indicate the
need for detailed magnetohydrodynamic modelling of GRB afterglows in
order to accurately determine the fireball parameters.


\section{Optical Decay}

\begin{figure}[htb]
   \begin{center}
      \epsfig{file=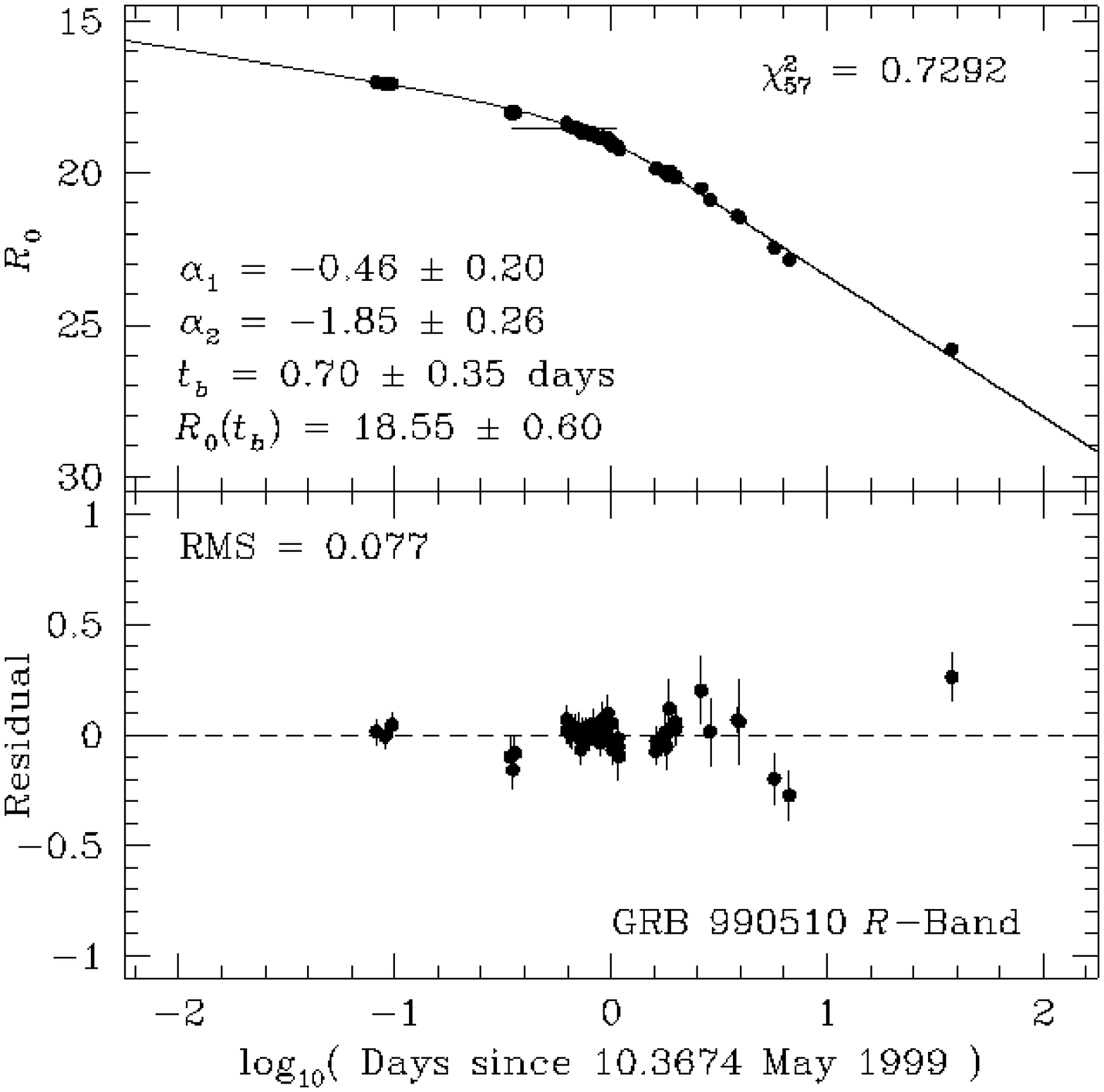,height=2.5in}
      \epsfig{file=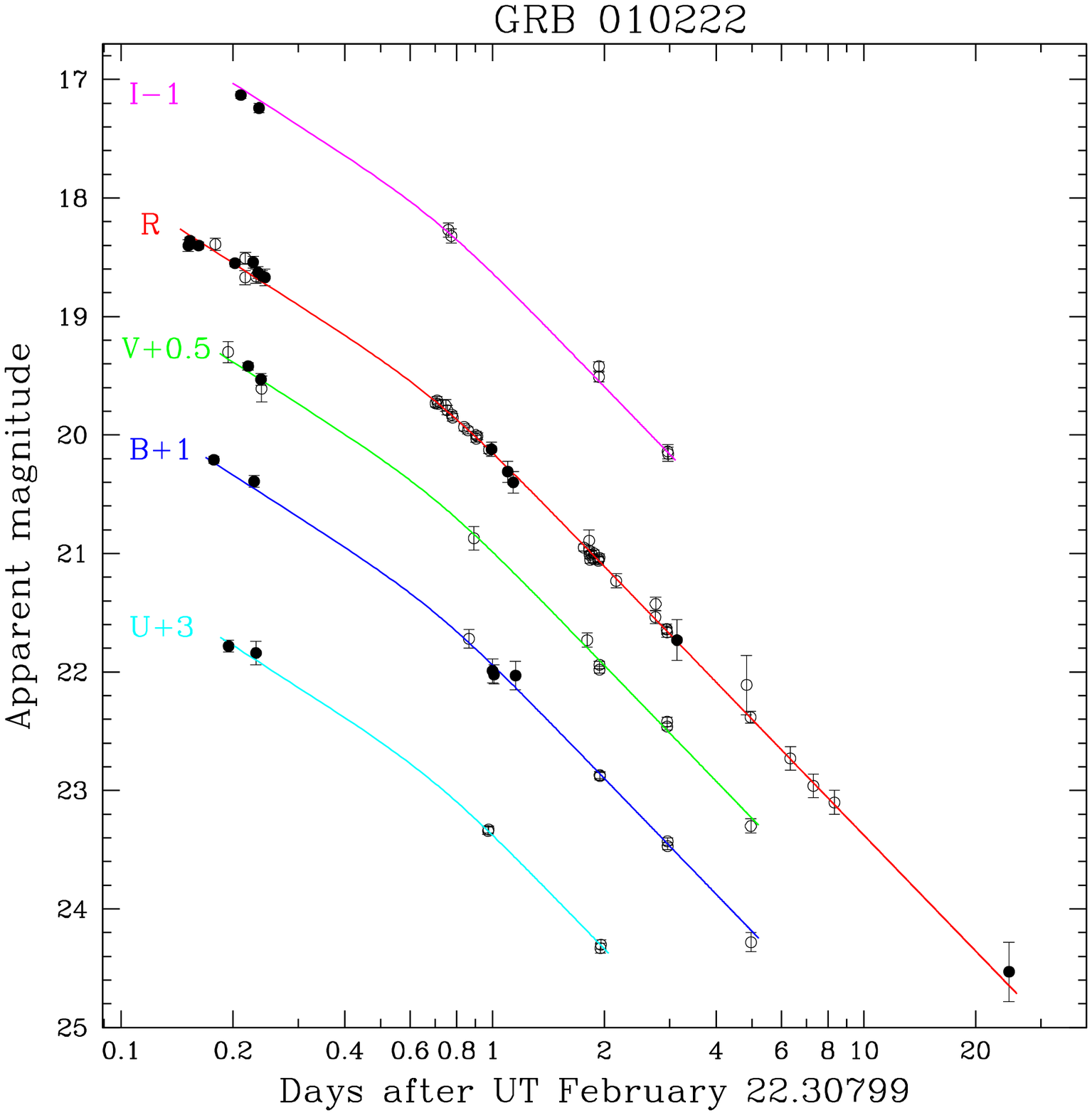,height=2.5in}
      \caption{The left Figure~\cite{HBH2000} shows the
               $R$-band decay for GRB~990510 (upper panel) and the
               residuals after subtracting the fitted optical decay
               (lower panel).  The right Figure shows the $U\!BV\!RI$
               optical decays for GRB~010222~\cite{SGJ2001}.  Note
               that the decays are achromatic, but that it is
               difficult to determine when the break
               occurs.\label{FIGURE:light_curves}}
   \end{center}
\end{figure}

     The rate of decay of the optical flux, $\alpha$, is related to
the value of the electron index.  These relationships are given
by~\cite{SPH1999} for an expanding sphere, and a jet, in a homogeneous
ambient medium.  Similarly,~\cite{CL1999} find corresponding
relationships for a burst in a pre-existing stellar wind.  If the
value of the electron index can be securely determined from spectra
then it can be used to predict the value of $\alpha$.  This can then
be compared to the observed optical decay to determine the relative
ordering of $\nu_c$ and $\nu_m$ with respect to the optical band, as
well as the nature of the ambient medium around the progenitor.  In
practice it is not always possible to determine $p$ solely from
spectral data, and sparse photometry can make it difficult to
determine the location of the break in the optical decay.  In these
cases $p$ must be determined simultaneously from the optical decay and
spectral slope using closure relations~\cite{PBR2002}.  However, it is
not always possible to unambiguously distinguish between different
cases.  Fig.~\ref{FIGURE:light_curves} shows examples of
well-determined optical decays.
 
\section{Breaks}

     There are several mechanisms that can cause a break in the
observed power-law decay of the optical afterglow of a GRB.  When the
cooling break passes through the optical (moving towards lower
frequencies in a homogeneous ambient medium and moving towards higher
frequencies in a pre-existing stellar wind) the slope of the optical
decay will change.  In a homogeneous ambient medium the decay becomes
steeper by $\Delta \alpha = 0.25$ while in a pre-existing stellar wind
the decay becomes shallower by the same amount.  The time of this
break depends on frequency, so it will occur at different times in
different optical bands.  A similar break will occur when the
synchrotron frequency passes through the optical.  This is expected to
occur within $\approx 2$ hours of the burst and the change in the
optical decay will depend on the details of the the ambient medium and
location of $\nu_c$ and $\nu_a$.

     Breaks have been seen in the optical decays of several
afterglows.  The first detections were~\cite{KDO1999}
and~\cite{CZC1999} who claimed a chromatic break in the optical decay
of GRB~990123.  The break was later shown to be achromatic
by~\cite{HBH2000}.  The first clear detection of an achromatic break
was by~\cite{SGK1999,HBH2000} (see Figure~\ref{FIGURE:light_curves}).
As of this writing breaks have been seen in 19 afterglows with break
times ranging from $\approx 0.5$~days for GRB~010222 to $\approx
25$~days for GRB~000418 and GRB~970508.  Since none of these breaks
have shown any dependence on colour it is unlikely that they are due
to a spectral break moving through the optical.

     Another break occurs when the relativistic fireball slows down
sufficiently that relativistic beaming stops.  The fireball is
initially expanding at highly relativistic speeds, so all of the
radiation from the external shock is relativistically beamed into a
cone with a half-opening angle of $\theta_b \approx 1 / \Gamma$ where
$\Gamma$ is the Lorentz factor of the expanding fireball.  As the
fireball expands into the surrounding medium it decelerates.  At early
times, when $\Gamma$ is large, all of the radiation is
relativistically beamed toward the observer.  However, when $\Gamma$
becomes small only that fraction of the light that is intrinsically
radiated towards the observer is visible.  This causes the observed
flux to decrease more rapidly and results in a break in the observed
decay rate.  The beaming break is a purely geometric effect and thus
will be achromatic.  The beaming break will occur regardless of the
geometry of the fireball and will result in the decay becoming steeper
by $\Delta \alpha = 0.75$.

     If the fireball is intrinsically collimated into a cone with a
half-opening angle of $\theta_j$ ({\ie}, a jet) then yet another break
can occur.  When $\Gamma$ drops to $\approx 1 / \theta_j$ the shock
front stops expanding and sideways expansion begins to dominate the
dynamics of the jet.  This leads to a rapid increase in the rate of
the optical decay, $\alpha_j$.  In general the late-time decay is
$\alpha_j \approx p$ once sideways expansion dominates.  The critical
Lorentz factors for both the onset of sideways expansion, and the end
of relativistic beaming are similar, so it is difficult to separate
these two breaks observationally.  In general the observed change in
the decay slope is significantly steeper than what is expected from
the end of relativistic beaming.  This strongly suggests that the
fireball is intrinsically collimated.  However, several bursts, such
as GRB~990510 and GRB~010222, exhibit breaks which appear to occur
over an extended period of time.  Some of this is due to sampling
uncertainties and some of it is probably a result of the processes
that cause the breaks occurring over a finite period of time.  In some
cases, however, the long durations may indicate that both beaming and
the sideways expansion are contributing to the change in the optical
decay at approximately the same time.


\section{Alternative Models}

     There are several alternate models to the standard relativistic
fireball for GRBs.  The cannonball model~\cite{DDD2002,DDD2003}
proposes that some supernov{\ae} eject ``cannonballs'' of baryonic
matter at highly relativistic speeds.  The gamma-ray pulses come from
interactions of the ejected material with circumstellar shells of
material.  These ``cannonballs'' are similar to those seen in
microquasars.  There are also several binary progenitor models which
propose that bursts are due to the merging of two compact objects.
These models include the binary neutron star model~\cite{NPP1992} and
the merger of a black hole and a neutron star~\cite{P1991}.


\section{The Late-Time Bump}

     To within the observational uncertainties the unusually faint, in
gamma rays, burst GRB~980425 occurred at the same time and location on
the sky as the unusually radio-loud Type Ib/c supernova
SN1998bw~\cite{GVV1998}.  This was the first evidence that some GRBs
might be related to supernov{\ae}.  The maximum brightness of a
supernova typically occurs approximately one week (in the rest frame)
after the explosion.  The optical afterglows of GRBs, on the other
hand, usually reach a maximum brightness within minutes or hours after
the burst.  GRB~970508 was an exception.  Its optical afterglow
remained at a constant luminosity for approximately one day then
brightened by a factor of about seven.  The difference in time scales
for the maximum brightness of supernov{\ae} and GRBs means that the
supernova component of the light should be visible as a rebrightening
over what is predicted from the power-law decay at $\approx 10(1+z)$
days after the burst.

     The first evidence for a late-time bump was
GRB~980326~\cite{BKD1999}.  The optical afterglow of this burst
brightened by $\approx 4$ mag over the magnitude predicted by the
best-fitting power law decay approximately 20 days after the burst.
This was interpreted by~\cite{BKD1999} as an overluminous SN~Ib/c at
$z \approx 1$ superimposed on the power-law decay of GRB~980326.
Since then bumps have been claimed for several bursts, but many of
these bumps have had low significance and are consistent with
uncertainties in the photometry or contamination from the host galaxy.

\begin{figure}[htb]
   \begin{center}
      \epsfig{file=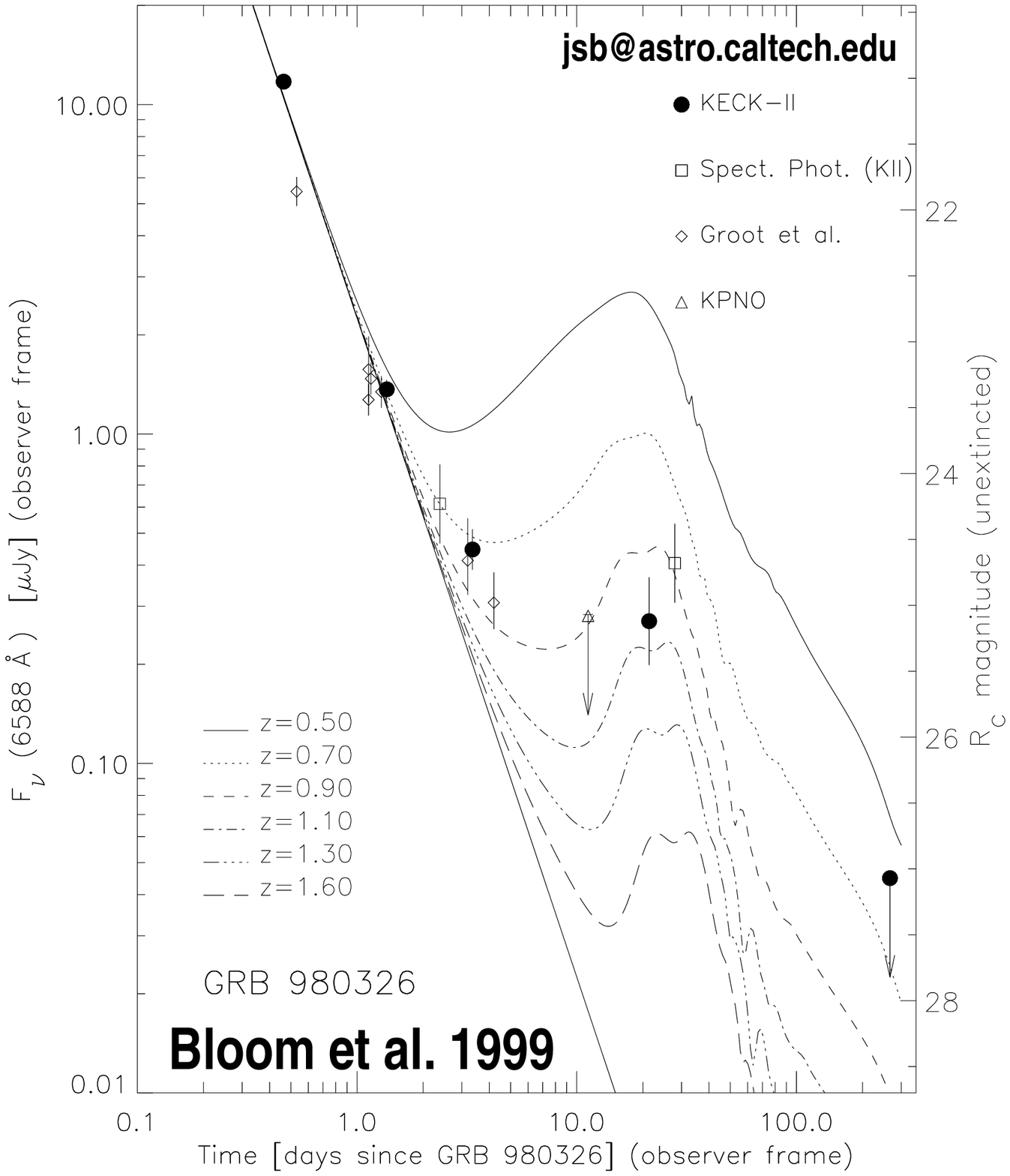,height=2.5in}
      \epsfig{file=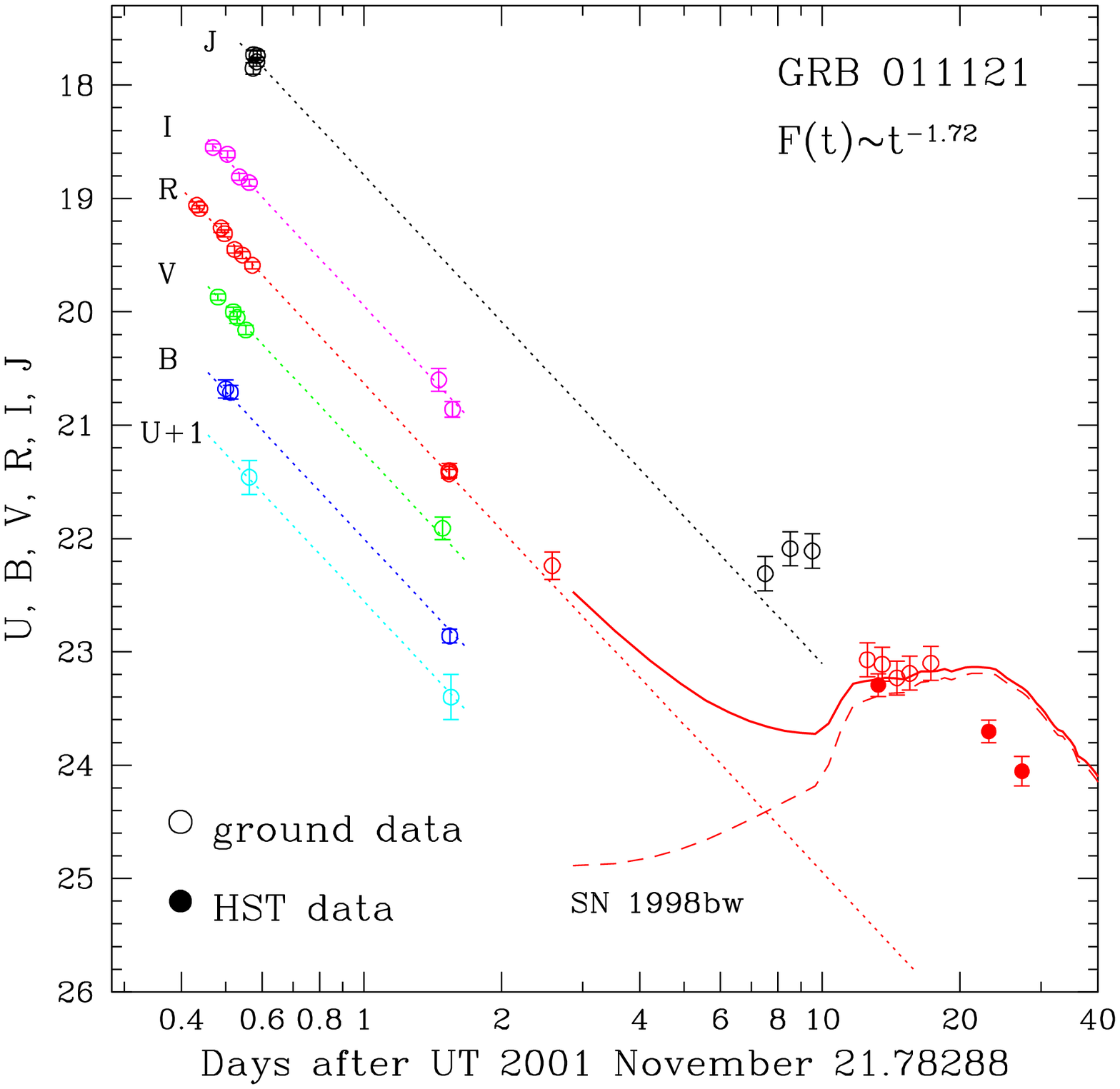,height=2.5in}
      \caption{The left figure~\cite{BKD1999} shows the bump for
               GRB~980326.  The right figure~\cite{GSW2002} shows the
               bump in the optical decay of GRB~011121 $\approx 20$
               days after the burst.\label{FIGURE:sn_bumps}}
   \end{center}
\end{figure}    

     The strongest evidence for an association between GRBs and
supernov{\ae} comes from optical observations of GRB~011121.  The
optical afterglow of this burst showed a large deviation from a
power-law decay starting approximately one week after the
burst~\cite{GSW2002,BKP2002}.  The duration and magnitude of this bump
were consistent with a supernova peaking $\approx 12$ days, in the
rest frame, after the burst.  The bump exhibited colour evolution that
was consistent with the local Type IIn supernova SN1998S.
Unfortunately the presence of a supernova component to GRB 011121 did
not become known until the {\sl HST\/}/WFPC2 images became public,
three months after they were obtained~\cite{GHJ2002}, so a golden
opportunity for intensive follow-up observations of the closest ($z =
0.362$) known classical GRB with a supernova component was lost.

     An alternate explanation for the late-time bump is light from the
burst being scattered by dust located between $\approx 0.1$ and 1.0~pc
from the progenitor.  The bumps seen in the optical decays of
GRB~970228 and GRB~980326 $\approx 20$--30 days after the burst are
consistent with such a dust echo~\cite{EB2000}.  As of this writing
there is no burst where a dust echo provides a better description of a
late-time bump than a supernova does.  Most late-time bumps can be
equally-well described by a supernova, a dust echo, or even no
bump~({\eg},~\cite{HFH2001}).


\section{Rapid variations}

\begin{figure}[htb]
   \begin{center}
      \epsfig{file=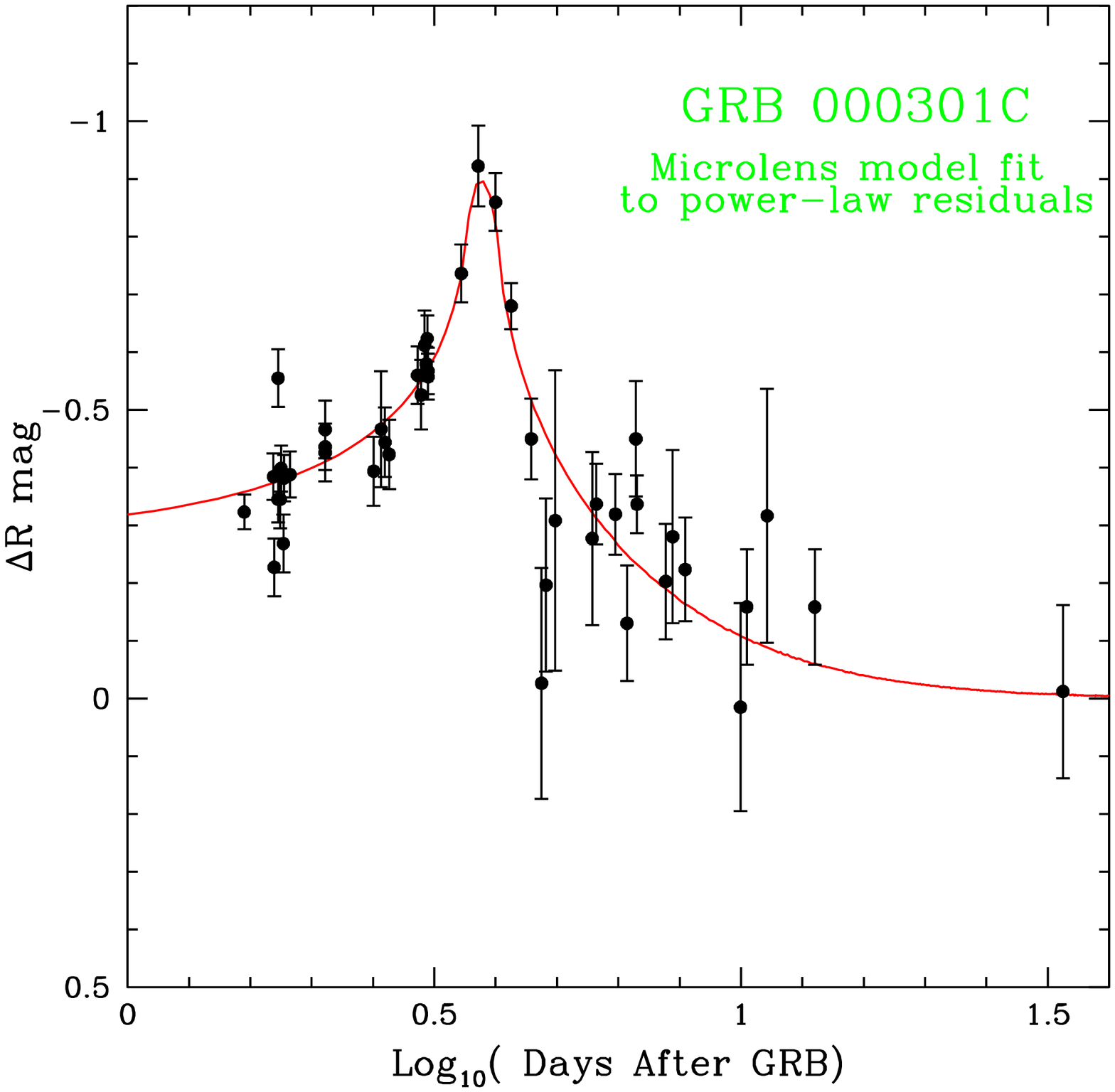,height=2.5in}
      \epsfig{file=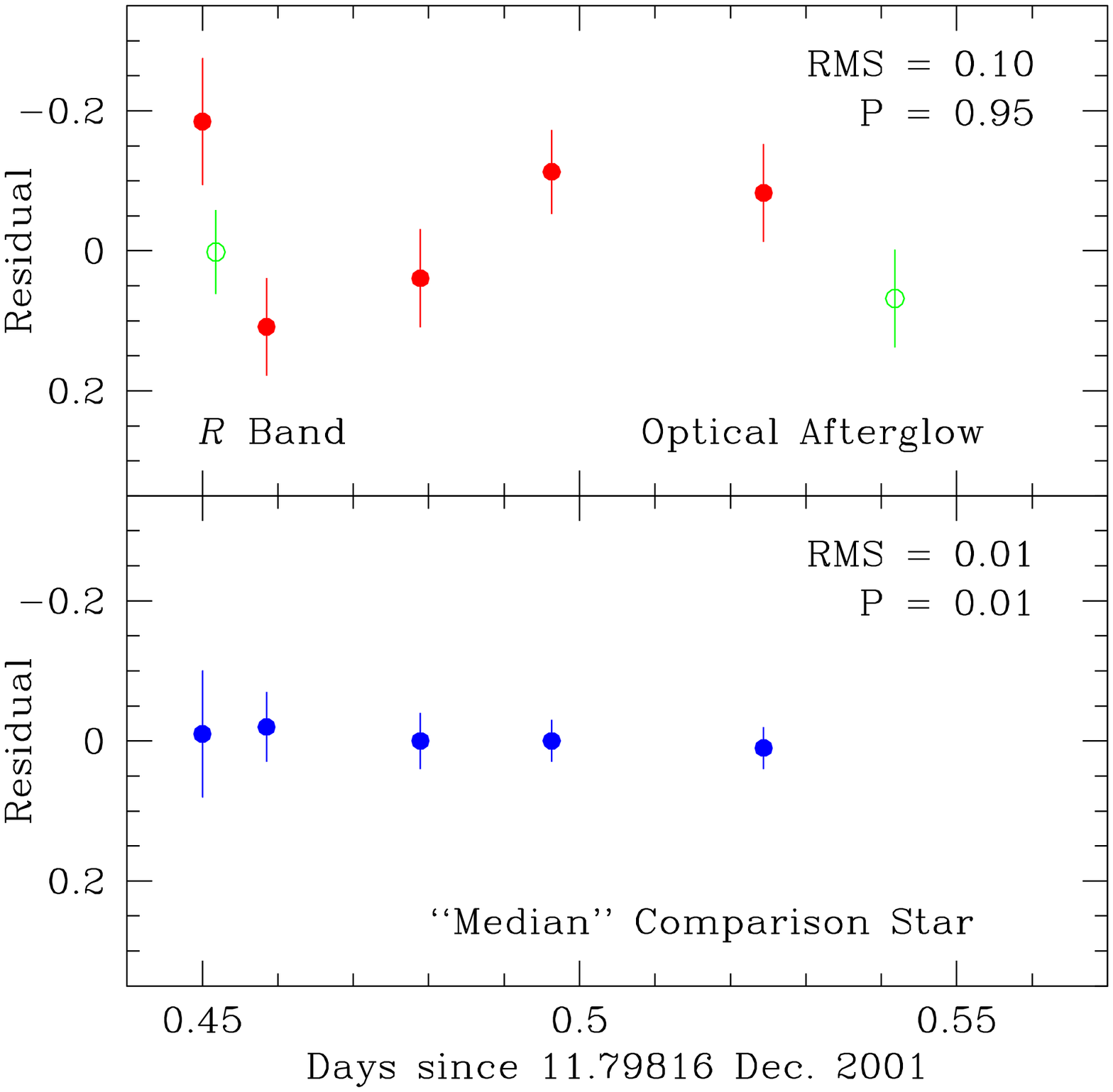,height=2.5in}
      \caption{The left figure~\cite{GLS2000} shows the $R$-band
               residuals for GRB~000301C after subtracting the
               best-fitting broken power law.  These residuals are
               well fit by gravitational lensing.  The right
               figure~\cite{HSG2002} shows the $R$-band photometry of
               GRB~011211 $\approx 0.5$ days after the burst shows
               rapid variations at the $\approx 8$\% level on
               time scales of $\approx$1--2
               hours.}
   \label{FIGURE:lc_vary}
   \end{center}
\end{figure}

     The classic GRB optical light decay shows no evidence for rapid
variations away from a power law decay.  As of the writing of this
article the two exceptions are GRB~000301C and GRB~011211 (see
Figure~\ref{FIGURE:lc_vary}).  GRB~000301C exhibited achromatic flux
variations of up to $\approx50$\% over time scales of several hours
between three and six days after the burst.  This was interpreted as
the signature of gravitational lensing from an approximately $0.5
\mathcal{M}_\sun$ star located at $z \approx 1$~\cite{GLS2000}.

     Rapid variations in the optical decay have been observed for
GRB~011211~\cite{HSG2002,J2002} approximately half a day after the
burst.  These variations are at the $\approx 8$\% level and have time
scales of $\approx 1$ hour.  They can be interpreted, using the
framework of~\cite{WL2000}, as being due to small-scale
inhomogeneities in the environment $\approx 0.05$--0.2 pc from the
burst's progenitor.  Their analyses of the optical light curve from
this burst suggests that GRB~011211 occurred in a circumburst medium
that is homogeneous over large scales but with density variations of a
factor of approximately two over distances of $\approx 40$--125 {\sc
AU}.

     These small-scale fluctuations are similar to those seen in the
interstellar medium in the Galaxy~\cite{DGR1989,FG2001}.  Similar
small-scale density fluctuations are also common in the environments
of Wolf--Rayet stars.  Radiative instabilities in Wolf--Rayet stars
can give rise to extensive structure and strong clumping in their
stellar wind~\cite{GO1995,H1991}.  If GRBs are related to the death
throes of massive stars then it is reasonable to expect similar
sub-structure in the local environments of GRBs.  This makes the
paucity of rapid fluctuations in most GRB light curves difficult to
explain.  One possible explanation is the geometry of the direction of
the progenitor's rotation axis relative to Earth.  Another possibility
is that the winds from the GRB progenitors are not symmetric.  If the
wind is denser near the equatorial regions of the progenitor, yet the
GRB occurs from the poles, the burst may avoid the bulk of the
structure in the local ambient medium.

\section{What about the future}

     The future of optical observations of GRBs promises to be an
exciting one.  The upcoming {\sl Swift\/} mission will dramatically
increase the rate of well-localized bursts.  An onboard
optical/ultraviolet telescope will be able to locate optical
afterglows to within a few arcseconds with-in minutes of the burst.
This rapid identification will permit study of the still-mysterious
early time behavior of the optical afterglow.  {\sl Swift\/}'s rapid
follow-up capabilities may also allow us to solve the puzzle of the
short--hard bursts.  Complementary to rapid follow-up observations
will be large-scale surveys and infrared observations.  These will
allow us to estimate the true rate of optical afterglows and thus
probe the nature of dark bursts.

     Perhaps the most important need for the future is continuous
monitoring of afterglows from immediately after the burst to as late
as possible.  We need to understand the early time behavior of
afterglows to disentangle the effects of internal, external, and
reverse shocks.  We also need to map out the breaks for as many bursts
as possible so that we can make statistically rigorous statements
about the times and durations of the breaks and their relation (if
any!) to the geometry of GRBs.  Finally, we need long-term monitoring
after each burst so that late-time bumps can be identified or ruled
out for all bursts.


\bigskip
This work was partially supported from the NASA LTSA grant NAG-9364,
by the Neils Bohr Institute for Astronomy, Physics, and Geophysics,
and by the University of Copenhagen.



\def\Discussion{
\setlength{\parskip}{0.3cm}\setlength{\parindent}{0.0cm}
     \bigskip\bigskip      {\Large {\bf Discussion}} \bigskip}
\def\speaker#1{{\bf #1:}\ }
\def\endDiscussion{}

\Discussion

\speaker{Edo Berger (Caltech)} Why wasn't a spectrum of GRB~011121
taken say after 14 days since it is the lowest redshift known GRB?\@
An indication of a SN from photometry was not really required.

     At 14 days after the burst there was no indication that a SN bump
was present.  In addition,~\cite{B2002} reported that {\sl HST\/}
observations showed ``no evidence of an intermediate-time light curve
bump (from an underlying supernova, etc.)''.  Telescope time is
valuable, so it was not used to obtain a spectrum of an object that
the initial analysis of the best available imaging data suggested did
not exist.  It was not until the {\sl HST\/} data became public and
was reanalysed by~\cite{GHJ2002} that there was any publicly-available
evidence for a supernova component.

\speaker{Edo Berger (Caltech)} In response to the speakers claim that
radio observations allow us to study events only out to $z \approx 1$,
I would like to point out that this is in fact incorrect ({\eg}
GRB~000301C).  The main problem is that both radio and optical probe a
narrow range of the broad-band synchrotron spectrum and therefore both
are needed to infer wind vs ISM circumburst medium.

     This is correct.  Radio can be used to study afterglows beyond $z
\approx 1$.

\speaker{S. Woosley (UCO/Lick)} One should be careful drawing
inferences pro or con regarding a SN presence based on the colour of
the bump in the optical afterglow of a GRB.\@ An aspherical SN viewed
down its explosion axis may have peculiar characteristics.  SN1998ba
accompanied a non-standard GRB (either because it had low explosion
energy or was viewed off axis.).  The colour of a SN Ic is sensitive
to the amount of Ni made.  Less Ni gives a blue supernova.
     
     This is a very good point.  What is needed is high-resolution
spectra of the late-time bumps which can be tested for supernova
signatures.

\speaker{Jens Hjorth (Copenhagen)} Do alternative models for the late
light curve bump explain the preference for bumps to peak at around
10--20 days at a peak magnitude comparable to a redshift SN1998bw
spectrum?

     There is no single explanation that I am aware of.

\speaker{D. Lazzati (Cambridge)} What is the increasing evidence that
bumps are related to SNe?  Aren't they simply consistent with SNe?

     All of the observed bumps are consistent with supernovae.  This
is not proof, but it is suggestive that there is a link.

\speaker{M. Rees (Cambridge)} The steepening in the light curves at
late times obviously fits nicely with the Rhoads model.  But if we
didn't believe in beaming for other reasons (energetics,
astrophysical, redshifting {\etc}), maybe we would not find it too
hard to interpret the data differently?

     I agree with this statement.  However, the Rhoads model
reproduces the general features seen in GRB afterglows.  This suggests
that the general features of the model are a reasonable approximation
to reality.

\speaker{M. Rees (Cambridge)} It would be rather surprising if the
spectral break at the cooling frequency were naturally very
sharp. There are various effects (reacceleration, inhomogeneities,
{\etc}) that could smear it out.  How sharp do the data require it to
be?

     The identification of break times is somewhat uncertain and can
depend on the fitting function used ({\eg},~\cite{HBH2000}).  The
times of the best-defined breaks tend to be uncertain by several
hours.  This is probably the smallest time scale that breaks occur on.
However, more bursts, and continuous monitoring of afterglows {\sl
during\/} their breaks are needed to address this issue properly.

\speaker{M. Rees (Cambridge)} Do you think SN1998bw could be a
``standard'' GRB that is misaligned without its radio emission being
stronger than is observed?

      I suspect that SN1998bw is one point on a continuum of GRBs.  At
present I lean towards the idea that SN1998bw was a standard,
collimated GRB which was oriented away from us.  However, I will leave
a a detailed response to those who are far more familiar with this
supernova than I am.

\endDiscussion
 
\end{document}